\documentclass[aps,prd,nofootinbib,showpacs,amssymb,twocolumn]{revtex4}
\usepackage{amsmath,amsfonts}
\usepackage{epsfig,subfigure}
\usepackage{ifpdf}
\usepackage{hyperref}
\usepackage{amsmath}
\usepackage{graphicx}
\usepackage{color}
\usepackage{natbib}
\usepackage{multirow}

\newcommand{\be}{\begin{equation}}
\newcommand{\ee}{\end{equation}}
\newcommand{\bea}{\begin{eqnarray}}
\newcommand{\eea}{\end{eqnarray}}

\newcommand\editremark[1]{ {\color{red} #1}}

\newcommand\hidetosubmit[1]{}
\renewcommand\hidetosubmit[1]{#1}
\newcommand\ForInternalReference[1]{}

\newcommand\unit[1]{\, {\rm #1}}
\newcommand\mc{ {{\cal M}_c}}

\newcommand\Y[1]{Y^{(#1)}}

\newcommand\qmstate[1]{\left|#1\right>}

\newcommand\qmstateproduct[2]{\left<#1|#2\right>}
\newcommand\qmoperatorelement[3]{\left<#1\left|#2\right|#3\right>}

\newcommand\NprecSpecific{7}

\newcommand\abbrvACST{ACST}
\newcommand\abbrvBLO{BLO}
\newcommand\abbrvLO{LO}
\newcommand\abbrvEFM{COOKL}
\newcommand\abbrvAlignedPE{OFOCKL}

\newcommand\citeMCMC{\cite{LIGO-CBC-S6-PE,2011PhRvD..83h2002D,2011PhRvD..84f2003C,gr-extensions-tests-Europeans2011,gwastro-mergers-PE-Aylott-LIGOATest,2011ApJ...739...99N,2012PhRvD..85j4045V,gw-astro-PE-Raymond,gw-astro-PE-lalinference-v1}}

\begin{document}

\author{Richard O'Shaughnessy$^{1}$}
\email{oshaughn@gravity.phys.uwm.edu}

\author{Benjamin Farr$^{2}$}

\author{Evan Ochsner$^{1}$}

\author{Hee-Suk Cho$^{3}$}

\author{V. Raymond$^{6}$}

\author{Chunglee Kim$^{4,5}$}

\author{Chang-Hwan Lee$^{5}$}

\affiliation{$^1$Center for Gravitation and Cosmology, University of Wisconsin-Milwaukee, Milwaukee, WI 53211, USA}

\affiliation{$^2$Department of Physics and Astronomy \& Center for Interdisciplinary Exploration and Research in
  Astrophysics (CIERA), Northwestern University, Evanston, IL, USA}

\affiliation{$^3$Department of Physics, Pusan National University, Busan 609-735, Korea}

\affiliation{$^4$Korea Institute of Science and Technology Information, Daejeon 305-806, Korea}

\affiliation{$^5$ Astronomy Program, Department of Physics and Astronomy, Seoul National University, 1 Gwanak-ro,
  Gwanak-gu, Seoul 151-742, Korea}

\affiliation{$^6$LIGO Laboratory, California Institute of Technology, MC 100-36, Pasadena CA, 91125, USA}

\title{Parameter Estimation of Gravitational Waves from Precessing BH-NS Inspirals with higher harmonics}
\date{\today}

\begin{abstract}
Precessing black hole-neutron star (BH-NS) binaries produce a rich gravitational wave signal, encoding the binary's nature and
inspiral kinematics.  
Using  the \texttt{lalinference\_mcmc} Markov-chain Monte Carlo parameter estimation code, we use two fiducial examples to
illustrate how the  geometry and kinematics are  encoded into the modulated gravitational wave signal,
using  coordinates well-adapted to precession.   
Extending previous work, we demonstrate the performance of detailed parameter estimation studies can often be estimated by
``effective'' studies: comparisons of a prototype signal with its nearest neighbors, adopting a fixed sky location and idealized two-detector network.   
Using a concrete example, we show higher harmonics provide nonzero but small
\emph{local} improvement when estimating the parameters of precessing BH-NS binaries.
We also show higher harmonics can improve parameter estimation accuracy for precessing binaries by breaking
leading-order discrete symmetries and thus ruling out approximately-degenerate source orientations.   
Our work illustrates quantities gravitational wave measurements can
provide, such as the  orientation of a precessing short gamma ray burst progenitor  relative to the line of
sight.  
More broadly,   ``effective'' estimates may provide a simple way to
\emph{estimate trends} in the performance of parameter estimation for generic precessing BH-NS binaries in
next-generation detectors.
For example,  our results suggest  that the orbital chirp rate, precession rate, and precession geometry  are roughly-independent observables, defining natural variables
to organize correlations in the high-dimensional BH-NS binary parameter space.
\end{abstract}

\maketitle

\section{Introduction}

Ground based gravitational wave detector networks (notably LIGO \cite{gw-detectors-LIGO-original-preferred} and Virgo
\cite{gw-detectors-VIRGO-original-preferred})  are sensitive to the relatively well understood signal from  the lowest-mass compact binaries
$M=m_1+m_2\le 16 M_\odot$ \cite{2003PhRvD..67j4025B,2004PhRvD..70j4003B,2004PhRvD..70f4028D,BCV:PTF,2005PhRvD..71b4039K,2005PhRvD..72h4027B,2006PhRvD..73l4012K,2007MNRAS.374..721T,2008PhRvD..78j4007H,gr-astro-eccentric-NR-2008,gw-astro-mergers-approximations-SpinningPNHigherHarmonics,gw-astro-PN-Comparison-AlessandraSathya2009}.  
Strong signals permit  high-precision constraints on binary parameters, particularly when the binary precesses.  
Precession arises only from spin-orbit misalignment; occurs on a distinctive timescale between the inspiral and orbit;
and produces distinctive polarization and phase modulations \cite{ACST,gw-astro-SpinAlignedLundgren-FragmentA-Theory,gwastro-SpinTaylorF2-2013}.  
As a result, the complicated gravitational wave signal from precessing binaries is unusually rich, allowing
high-precision constraints on multiple parameters, notably the (misaligned) spin \cite{LIGO-CBC-S6-PE,gwastro-mergers-HeeSuk-FisherMatrixWithAmplitudeCorrections}.  
Measurements of the spin orientations alone could provide insight into processes that misalign spins and orbits, such as
supernova kicks \cite{2013MNRAS.434.1355J,2012MNRAS.423.1805N}, or realign them, such as tides and post-Newtonian resonances
\cite{2013PhRvD..87j4028G}.  
More broadly, gravitational waves constrain the pre-merger orbital plane and total angular momentum direction, both of
which may correlate with the presence, beaming, and light curve \cite{2010ApJ...722..235V,2011ApJ...733L..37V,2013ApJ...767..141V} of any post-merger ultrarelativistic blastwave (e.g, short GRB)
 \cite{2009ARAA..47..567G}.  
Moreover, spin-orbit coupling strongly influences orbital decay and hence the overall gravitational wave phase: the
accuracy with which most other parameters can be determined is limited by knowledge of BH spins
\cite{1995PhRvD..52..848P,2013PhRvD..87b4035B,gwastro-mergers-HeeSuk-FisherMatrixWithAmplitudeCorrections,gwastro-mergers-HeeSuk-CompareToPE-Aligned}.  
Precession is known to break this degeneracy
\cite{2006PhRvD..74l2001L,2009PhRvD..80f4027K,2011PhRvD..84b2002L,gwastro-mergers-HeeSuk-FisherMatrixWithAmplitudeCorrections,2007CQGra..24..155V,LIGO-CBC-S6-PE}.  
In sum,  the rich  gravitational waves emitted from a precessing binary  allow higher-precision measurements of
individual neutron star masses, black hole masses,  and and black hole spins, enabling  constraints on their distribution across multiple events.  In
conjunction with electromagnetic measurements, the complexity of a fully precessing gravitational wave signal may enable
correlated electromagnetic and gravitational wave measurements to much more tightly constrain the central engine of
short gamma ray bursts.

Interpreting gravitational wave data requires systematically comparing all possible candidate signals to the data,
constructing a Bayesian posterior probability distribution for candidate binary parameters \citeMCMC{}.   Owing to the
complexity and multimodality of these posteriors, successful strategies adopt two elements: a well-tested generic
algorithm for parameter estimation, such as variants of Markov Chain Monte Carlo or nested sampling; and deep insight into the structure of
possible gravitational wave signals, to ensure efficient and complete coverage of all possible options \cite{gw-astro-PE-Raymond-JumpProposals,gw-astro-PE-systemframe}.
Owing both to the relatively large number of parameters needed to specify a precessing binary's orbit and to the
seemingly-complicated evolution,   Bayesian parameter estimation methods have only recently  able to efficiently draw
inferences about gravitational waves from precessing sources \cite{gw-astro-PE-systemframe}.  
These improvements mirror and draw upon a greater theoretical appreciation of the surprisingly simple dynamics and
gravitational waves from precessing binaries, both 
in the post-Newtonian limit
\cite{ACST,2012PhRvD..86h4017B,gwastro-mergers-nr-Alignment-ROS-PN,gwastro-SpinTaylorF2-2013,2013PhRvD..88l4015K}  and strong
field
\cite{gwastro-mergers-nr-Alignment-ROS-Polarization,gwastro-mergers-nr-Alignment-ROS-CorotatingWaveforms,gwastro-mergers-nr-Alignment-BoyleHarald-2011,gwastro-mergers-nr-ComovingFrameExpansion-TransitionalHybrid-Schmidt2012,gwastro-mergers-nr-ComovingFrameExpansionSchmidt2010,gwastro-nr-imrphenomP}.  
For our purposes, these insights have suggested particularly well-adapted coordinates with which to express the dynamics
and gravitational waves from precessing BH-NS binaries, enabling more efficient and easily understood calculations.  
In particular, these coordinates have been previously applied to estimate how well BH-NS parameters can be measured by
ground-based detectors \cite{gwastro-mergers-HeeSuk-FisherMatrixWithAmplitudeCorrections}.  
In this work, we will present the first detailed parameter estimation calculations which  fully benefit from these
insights into  precessing dynamics.
In short, we will review the  natural parameters to describe the gravitational wave signal; demonstrate how well they
can be measured, for a handful of selected examples; and interpret our posteriors using simple, easily-generalized
analytic and geometric arguments.

As a concrete objective, following prior work
\cite{gwastro-mergers-HeeSuk-FisherMatrixWithAmplitudeCorrections,gwastro-mergers-HeeSuk-CompareToPE-Aligned} we will
explore whether higher harmonics break degeneracies and provide additional information about black hole-neutron star
binaries.  In the absence of precession, higher harmonics
are known to break degeneracies and improve sky localization, particularly for LISA
\cite{2006PhRvD..74l2001L,2009PhRvD..80f4027K,2011PhRvD..84b2002L}.   That said, these and other studies also suggest
that higher harmonics provide relatively little additional information about generic precessing
binaries, over and above the leading-order quadrupole radiation
\cite{gwastro-mergers-HeeSuk-FisherMatrixWithAmplitudeCorrections,2011PhRvD..84b2002L}.
For example, for  two fiducial nonprecessing and two fiducial precessing signals,
\citet{gwastro-mergers-HeeSuk-FisherMatrixWithAmplitudeCorrections}, henceforth denoted \abbrvEFM{}, provide concrete predictions for how well detailed parameter estimation
strategies should perform, for a specific waveform model.  
A previous work \cite{gwastro-mergers-HeeSuk-CompareToPE-Aligned}, henceforth denoted \abbrvAlignedPE{}, demonstrated these simple predictions accurately
reproduced the results of detailed parameter estimation strategies. 
In this work, we report on detailed parameter estimation for the two fiducial precessing signals described in
\abbrvEFM{}.  
As with nonprecessing binaries, we find higher harmonics seem to provide significant insight into geometric
parameters, in this case the projection of the orbital angular momentum direction on the plane of the sky.   
As this orientation could conceivably correlate with properties of associated electromagnetic counterparts, higher harmonics may
have a nontrivial role in the interpretation of coordinated electromagnetic and gravitational wave observations. 

This paper is organized as follows.  
In Section \ref{sec:Review} we describe the gravitational wave signal from precessing BH-NS binaries, emphasizing
suitable coordinates for the spins  (i.e., defined at $100\unit{Hz}$, relative to the total angular momentum direction) and
the waveform (i.e., exploiting the corotating frame to decompose the signal into three timescales: orbit,
precession, and inspiral).  
Our description of gravitational waves from precessing BH-NS binaries follows
\citet{gw-astro-SpinAlignedLundgren-FragmentA-Theory}, henceforth denoted \abbrvBLO, and \cite{gwastro-SpinTaylorF2-2013},
henceforth denoted \abbrvLO.  
Next, in Section \ref{sec:Results} we describe how we created synthetic data consistent with the two fiducial precessing
signals described in \abbrvEFM{} in gaussian noise;   reconstructed a best estimate (``posterior distribution'') for the possible
precessing source parameters consistent with that signal; and compared those predictions with semianalytic estimates.  
These semianalytic estimates generalize work by \abbrvEFM{}, approximating the full response of a multidetector network with a
simpler but more easily understood expression.  
Using simple analytic arguments, we describe how to reproduce our full numerical and semianalytic results using a simple separation of scales and physics: orbital cycles, precession cycles, and
geometry.   The success of these arguments can be extrapolated to regimes well outside its
limited scope, allowing simple predictions for the performance of precessing parameter estimation.  
We conclude in Section \ref{sec:Conclusions}.

For the benefit of experts, in 
Appendix \ref{ap:RevisedEffectiveFisher} we discuss the numerical stability and separability of our effective Fisher
matrix.  

\section{Kinematics and gravitational waves from precessing BH-NS binaries}
\label{sec:Review}

\begin{figure}[t!]
\includegraphics[width=8cm,height=6cm]{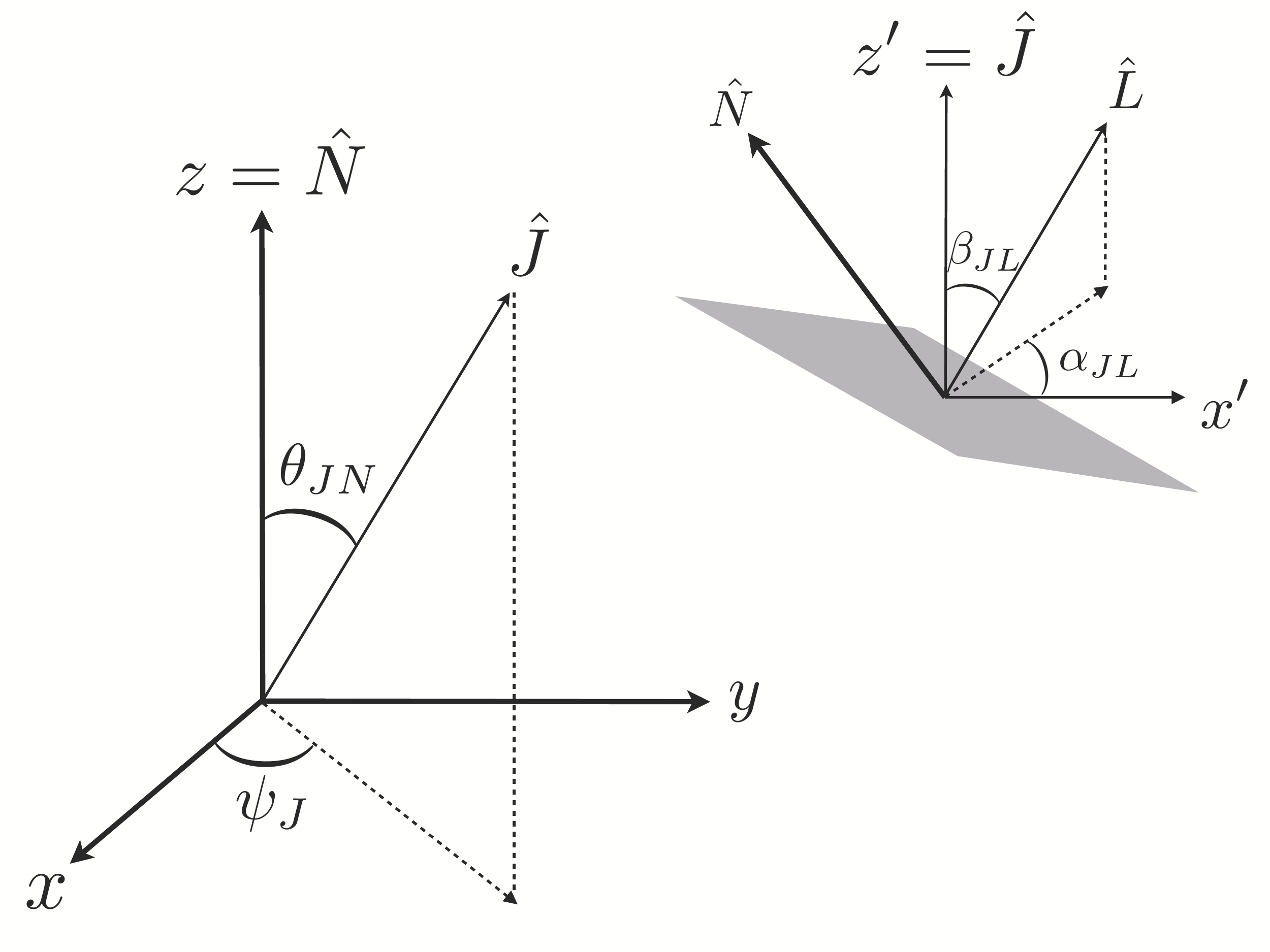}
\caption{\label{fig:Coords}{\bf Coordinate system for the precessing binary}. The left coordinate corresponds to the conventional GW radiation frame. $\theta_{JN}$ ($\phi_{JN}$) is a polar (azimuthal) angle of the total angular momentum ($J$) with respect to the radiation vector ($N$). In the right coordinate $\beta_{JL}$ ($\alpha_{JL}$) is a polar (azimuthal) angle of the orbital angular momentum ($L$) with respect to the total angular momentum ($J$). In the right coordinate, $N$, $J$, and ${\rm x'}$ are coplanar and the shaded region indicates the orbital plane.\label{fig1}} 
\end{figure}

\begin{table*}[!]
\begin{tabular}{c|ccccccccccc|c}
Name&$m_1$&$m_2$& $\chi_1$ & $\psi_J$ &$\psi_L$ &$\iota$ &$\beta_{JL}$ &  $\theta_{JN}$ 
& $\alpha_{JL}$ & $\theta_{LS_1}$ & $\phi_{\rm ref}$&$f_{\rm MECO}$  \\ 
  & $M_\odot$ & $M_\odot$ &  & & & & & & & & & Hz
\\
\hline
A&10&1.4&1.0& 1.25 & 0.72   & 1.512 &  $\pi/4$ & 0.730 & 2.95 &1.176   & 0.93 &  810    \\  %
C&10&1.4&1.0& 1.09 & 2.23   & 0.411 & $\pi/4$ & 0.891 & 5.75  & 1.176 &  1.65 & 810   \\
\end{tabular}
\caption{\label{tab1}{\bf Fiducial source parameters for precessing binaries.} We adopt  two fiducial binaries
  A,  C \emph{similar} to those used in \abbrvEFM{}.   All parameters are specified when twice the orbital frequency
  is $100 \unit{Hz}$. 
The post-Newtonian signals used in the text terminate at an orbital  $f_{\rm MECO}/2$, where $f_{\rm MECO}$ is the smaller of the  ``minimum
energy circular orbit'' (hence the acronym) and the frequency at which $\dot{\omega}<0$; the values shown are derived
from the same \texttt{lalsimulation} output used in our simulations, estimated from data evaluated at a $32\; kHz$ sampling
rate. %
Comparing with model waveforms that include inspiral,merger, and ringdown, we anticipate this abrupt termination causes
relatively little mismatch between our model and the physical signal; see \abbrvAlignedPE{}.  
}
\end{table*}

\begin{table}
\begin{tabular}{llllll}
$d$ & $t$ & $DEC$ & $RA$ & $\Delta t_{LH} $ & $\Delta t_{VH} $ \\
Mpc & s &  &  & ms & ms  \\
\hline
  23.1  &
 894383679.0 &
0.5747   &
  0.6485 &
  $- 3.93 $ &
 $ 5.98 $ \\
\end{tabular}
\caption{\label{tab:SourceGeometryOnSky}\textbf{Source location}: Source geocenter event time and sky location.  For a
  sense of scale, this table also provides  the
  time differences between different detector sites,  implied by that sky location and event time.
}
\end{table}

\begin{figure}[!]
\includegraphics[width=\columnwidth]{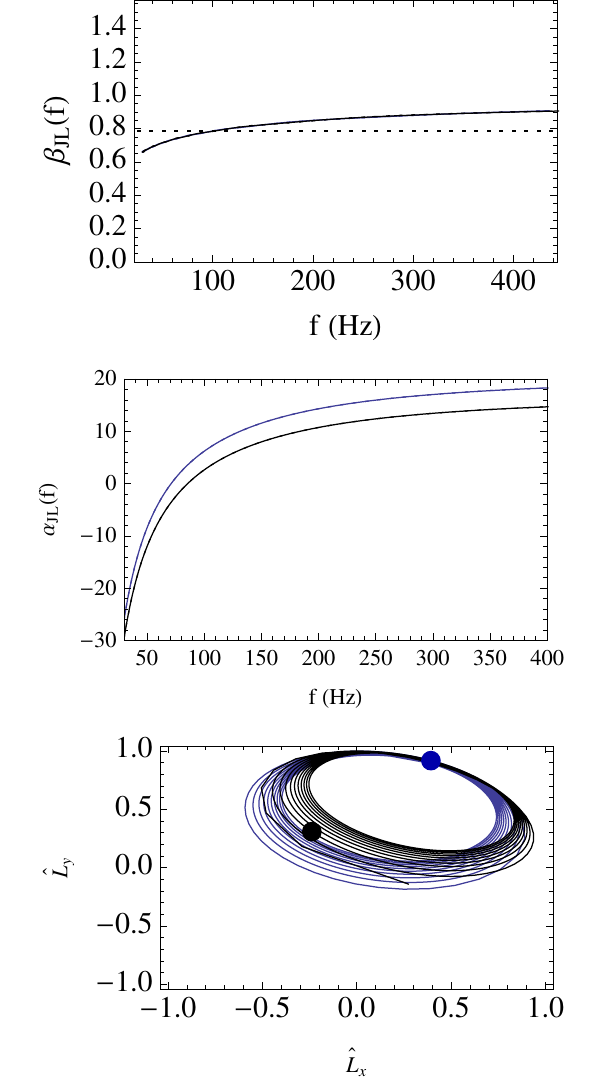}
\caption{\label{fig:AnglesVersusTime}\textbf{Time-dependent geometry}: For each of the two fiducial binaries in Table
  \ref{tab1},  a plot of the time-dependent angles $\beta_{JL}(t)$ [top; common to both],  $\alpha(t)$ plus an arbitrary
      integer multiple of $2 \pi $ [center;
    blue=A; %
 black=C], and the two components of the orbital angular momentum direction  $\hat{L}$ projected on the plane of the sky [bottom].   
    For completeness, the first two panels also include dotted lines, corresponding to the trajectory adopted when
    higher harmonics were included and the initial template frequency was reduced.  
    In the bottom  panel, solid colored dots indicate the direction of $\hat{L}$ at $100 \unit{Hz}$. 
The binary precesses roughly \NprecSpecific{} times between $30\unit{Hz}$ and $500\unit{Hz}$
}
\end{figure}

\subsection{Kinematics and dynamics of  precessing binaries }

The kinematics of precessing binaries are well described in \cite{ACST}, \abbrvBLO{}, and \abbrvLO; see, e.g., Eq. (10)
in \abbrvBLO.  In brief, the orbit contracts in the
instantaneous orbital plane on a long timescale $1/\Omega_{rad}$ over many orbital periods $1/\Omega_\phi$.  On an
intermediate timescale $1/\Omega_{prec}$,  due to spin-orbit coupling the angular momenta precess around the total angular momentum
direction, which remains nearly constant.  On timescales $1/\Omega_{\rm prec}$ between $1/\Omega_{\phi}$ and $1/\Omega_{rad}$, the orbital
angular momentum traces out a ``precession cone''.   
For this reason, we adopt coordinates at $100 \unit{Hz}$ which describe the orientation of all angular momenta relative
to $\vec{J}$; our coordinates are
identical to those used in \abbrvBLO, \abbrvLO, and \abbrvEFM{}.   Relative to a frame with $\hat{z}$ oriented along the
line of sight, the total, orbital, and spin  angular momenta are described by the vectors:\footnote{Strictly
  speaking, the total angular momentum $\vec{J}$ precesses \cite{ACST}.  For the results described in this work, we
  always adopt the total angular momentum direction evaluated at $f=100 \unit{Hz}$.}
\begin{align}
\hat{J}  &= \sin \theta_{JN} \cos \psi_J \hat{x} + \sin \theta_{JN} \sin \psi_J \hat{y} + \cos \theta_{JN} \hat{z}  \\
\hat{L}  & = \sin \iota \cos \psi_L \hat{x} + \sin \iota \sin \psi_L \hat{y} + \cos \iota \hat{z}  \\
\hat{S}_1  & = \sin \theta_1 \cos (\psi_L+\phi_1) \hat{x} + \sin \theta_1 \sin( \psi_L+\phi_1) \hat{y} + \cos \theta_1 \hat{z}  
\end{align}
where in this and subsequent expressions we restrict to a binary with a single spin (i.e., $\vec{S}_2=0$).  
Because the orbital angular momentum evolves along a cone, precessing around $\hat{J}$, we prefer to describe the orbital and spin angular momenta in  frame aligned with the total angular momentum $\hat{z}'=\hat{J}$:
\begin{eqnarray}
\hat{L} =  \sin \beta_{JL} \cos \alpha_{JL} \hat{x}' + \sin \beta_{JL} \sin \alpha_{JL} \hat{y}' + \cos \beta_{JL} \hat{J} 
\end{eqnarray}
where the frame is defined so  $\hat{y}'$ is perpendicular to $\hat{N}$ as in Figure \ref{fig:Coords}
\cite{2003PhRvD..67j4025B}: 
\begin{align}
\hat{y}' &= -  \frac{\hat{N}\times \hat{J}}{|\hat{N}\times \hat{J}|} 
\;,  \quad \;
\hat{x}' =  \hat{y}'\times \hat{J} =  \frac{\hat{N} - \hat{J}(\hat{J}\cdot \hat{N})}{|\hat{N}\times \hat{J}|}
\end{align}
In this phase convention for  $\alpha_{JL}$, the zero of $\alpha_{JL}$ is one of the two points when $\hat{L},\hat{J},\hat{N}$ are
all in a common plane,
sharing a common direction in the plane of the sky. %
Transforming between these two representations for $\hat{L}$ is straightforward.  For example, given  $\hat{N},\hat{L}$ and $\hat{J}$, we identify $\alpha$ and $\beta_{JL}$ via
\begin{align}
\beta_{JL}  &= \cos^{-1} \hat{J} \cdot \hat{L} \\
\alpha_{JL}& =\text{arg} \hat{J}\cdot[ \frac{\hat{L} \times (\hat{x}'+ i \hat{y}')}{i\sin\beta_{JL}}]  %
\end{align}
The spin angular momentum direction is determined from the direction of $\hat{L}$, the direction of $\hat{J}$, and the
angle $\theta_{LS}$ between $\hat{S}_1$ and $\hat{L}$:
\begin{align}
\hat{S}_1 &=  \sin (\beta_{JL}-\theta_{LS}) \cos \alpha_{JL} \hat{x}' + \sin (\beta_{JL}-\theta_{LS}) \sin \alpha_{JL} \hat{y}'
\nonumber \\
 &  + \cos (\beta_{JL}-\theta_{LS}) \hat{J} 
\end{align}
Finally, the opening angle $\beta_{JL}$ and the angle $\theta_{LS}$ are related.
Using  the ratio of $\vec{S}_1$ to the Newtonian angular momentum $\vec{L}=\mu \vec{r}\times \vec{v}$ as a parameter:
\begin{eqnarray}
\label{eq:def:gamma}
\gamma(t) &\equiv& |\vec{S}_1|/|\vec{L}(t)| = \frac{\chi_1 m_1^2}{\eta M\sqrt{M r(t)} } = \frac{m_1 \chi_1}{m_2} v
\end{eqnarray}
Using this parameter, the opening angle  $\beta_{JL}$ of the precession cone (denoted $\lambda_L$ in \abbrvACST) can be expressed trigonometrically as 
\begin{eqnarray}
\label{eq:def:SingleSpin:beta:Evolve}
\beta_{JL}(t)& \equiv& \arccos \hat{J}\cdot \hat{L} 
 =  \arccos \frac{1+\kappa \gamma}{\sqrt{1+2\kappa \gamma +\gamma^2}}
\end{eqnarray}
where $\kappa = \cos \theta_{LS} = \hat{L}\cdot \hat{S}_1$.  
Most BH-NS binaries' angular momenta evolve via simple precession:  $\alpha$ increases nearly uniformly on the
precession timescale, producing several precession cycles in band [Eq. (9) in \abbrvBLO{}], while $\beta_{JL}$ increases
slowly on the inspiral timescale, changing opening angle only slightly [Fig. 1 in \abbrvBLO{}]; see Figure \ref{fig:AnglesVersusTime}.

As
described  in
\abbrvAlignedPE{}, we evolve the angular momenta evolve according to expressions derived from general relativity in the
post-Newtonian,
adiabatic, orbit-averaged limit, an approximation presented in \cite{ACST} and described \cite{2004PhRvD..70l4020S}.  
Though some literature adopts a purely Hamiltonian approach to characterize spin precession
\cite{gwastro-approx-EOB-DamourEtAl2008-NextToLeadingOrderSpins,2008PhRvD..77j4018S,2011CQGra..28q5001L,2011PhRvD..84h4028N,Hartung:2013dza,2013PhRvD..87f4035T},
this orbit-averaged  approach is usually adopted when simulating gravitational waves from precessing binaries \cite{BCV:PTF,gw-astro-mergers-approximations-SpinningPNHigherHarmonics,2013PhRvD..87j4028G}.
In this work we adopt two fiducial precessing BH-NS binaries, with  intrinsic  and extrinsic parameters specified in
Tables \ref{tab1} and \ref{tab:SourceGeometryOnSky}.   Figure \ref{fig:AnglesVersusTime} shows how each binary precesses
around the total angular momentum direction $\hat{J}$ and in the plane of the sky.  
For this mass ratio, the opening angle $\beta_{JL}$ adopted in consistent with randomly-oriented BH spin; see, e.g.,
Eq. (\ref{eq:def:SingleSpin:beta:Evolve}) and Fig. 4 of \abbrvLO.
For this sky location, our simplified three-detector gravitational wave network has comparable sensitivity to both linear (or both
circular) polarizations.  

\subsection{Gravitational waves from precessing binaries}
Precession introduces modulations onto the ``carrier signal'' produced by the secular decay of the orbit over time.  \abbrvBLO{}
and \abbrvLO{}   provide a compact summary of the associated signal, in the time and frequency domain.  
In a frame aligned with the total angular momentum, several harmonics $h_{lm}$  are significant:
\begin{eqnarray}
h_+ - i h_\times = \sum_{lm} h_{lm} \Y{-2}_{lm}
\end{eqnarray}
where the harmonics $h_{lm}$ are provided and described in the literature \cite{gw-astro-mergers-approximations-SpinningPNHigherHarmonics}.
By ``significant'',  we mean that harmonics have nontrivial power $\rho_{lm}$:
\begin{eqnarray}
\rho_{lm}^2 = 2\int_{-\infty}^{\infty} \frac{|\tilde{h}_{lm}|^2}{S_h(f)}
\end{eqnarray}
where $S_h$ is the fiducial initial LIGO design noise power spectrum.  
These precession-induced modulations are most easily understood in a corotating frame, as in \abbrvLO
\cite{gwastro-mergers-nr-Alignment-ROS-Polarization,gwastro-mergers-nr-Alignment-ROS-Methods,gwastro-mergers-nr-Alignment-ROS-PN,gwastro-mergers-nr-Alignment-BoyleHarald-2011,gwastro-mergers-nr-ComovingFrameExpansion-TransitionalHybrid-Schmidt2012,gwastro-mergers-nr-ComovingFrameExpansionSchmidt2010,gwastro-nr-imrphenomP}:
\begin{eqnarray}
\label{eq:CorotatingFrame}
h_{lm} = \sum_{m'}D^l_{mm'}(\alpha_{JL},\beta_{JL},\gamma)h_{lm'}^{\rm ROT}
\end{eqnarray}  
where $\gamma = -\int d\alpha \cos \beta_{JL}$ and where $D^l_{mm'}$ is a Wigner D-matrix.  In this expression,  $h_{lm}^{\rm ROT}$ is the gravitational wave signal emitted by a binary with instantaneous angular momentum along
the $\hat{L}$ axis.
In the low-velocity limit, $h_{lm}^{\rm ROT}$ is dominated by leading-order radiation and hence by equal-magnitude $(l,m)=(2,\pm 2)$
modes.  Due to spin-orbit precession with $\beta_{JL} \ne 0$, however,   these harmonics are mixed.  When $\beta_{JL}$ is greater
than tens of degrees, then in the simulation frame
all $h_{lm}$ are generally present and significant.  

\begin{figure}
\includegraphics[width=\columnwidth]{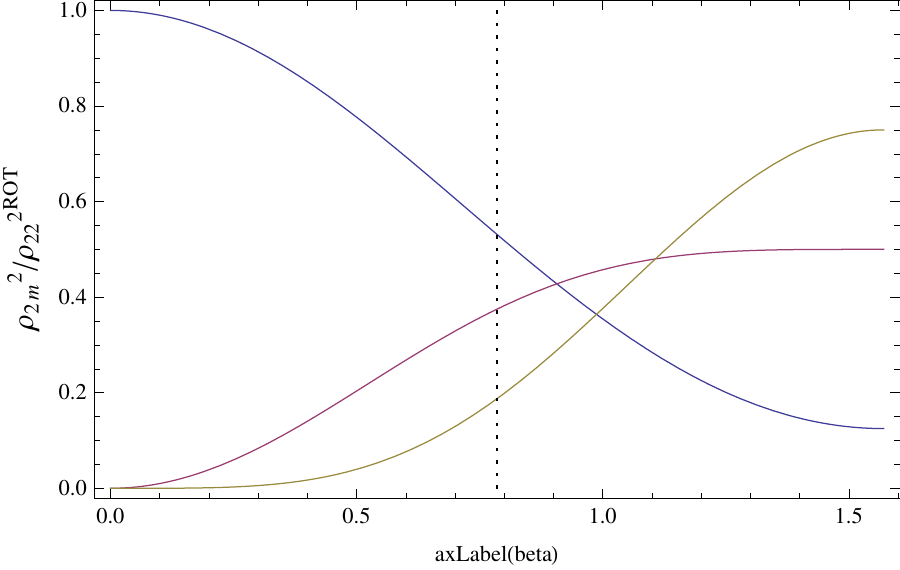}
\caption{\label{fig:IllustrateHarmonicPowerVersusBeta}\textbf{Harmonic amplitude versus opening angle}: A plot of
  $\rho_{2m}^2/(\rho_{22}^{\rm ROT})^2$ predicted by Eq. (\ref{eq:RhoEstimateLeading}) for $m=2$ (blue), $1$ (red), and
  $0$ (yellow).   
}
\end{figure}

To illustrate that gravitational wave emission from a precessing binary requires several harmonics $h_{lm}$ to describe
it when $\beta_{JL} >0$,  we evaluate $\rho_{2m}$, conservatively assuming only the $(2,\pm 2)$ corotating-frame modes are nonzero:
\begin{align}
\rho_{2m}^2 &\simeq [\rho_{2,2}^{\rm ROT}]^2 |d^2_{2,m}(\beta_{JL})|^2 + [\rho_{2,-2}^{\rm ROT}]^2 |d^2_{-2,m}(\beta_{JL})|^2
\nonumber \\
& =\rho_{2,2}^{\rm ROT} [|d^2_{2,m}(\beta_{JL})|^2 + |d^2_{-2,m}(\beta_{JL})|^2]
\label{eq:RhoEstimateLeading}
\end{align}
where we use orthogonality of the corotating-frame $(2,\pm 2)$ modes.    Figure
\ref{fig:IllustrateHarmonicPowerVersusBeta} shows that except for a small region $\beta_{JL} \simeq 0$, several harmonics 
contribute significantly to the amplitude along generic lines of sight, with $\rho_{2m}/\rho_{22}\gtrsim 0.1$.  
At this level, these harmonics change the signal significantly, both in overall amplitude ($\rho^2/2 * 0.1^2 \simeq
\rho^2 0.05$) and in fit to candidate data.
Gravitational waves from precessing BH-NS binaries are modulated  in amplitude, phase, and polarization.    A generic precessing source oscillates between emitting preferentially right-handed and preferentially left-handed
radiation along any line of sight; see \cite{gwastro-mergers-nr-Alignment-ROS-Polarization}.  
For the scenario adopted here, however, the orbital angular momentum almost always preferentially points towards the
observer ($\hat{L}\cdot \hat{N}\gtrsim 0$), so gravitational waves emitted along the line of sight are 
principally right handed for almost all time. %

\subsection{Symmetry and degeneracy}
Gravitational waves are spin 2: the spin-weight $-2$ expression $h=h_+-i h_\times$ transforms as $h\rightarrow h \exp
(-2i\psi)$ under a rotation by $\psi$ around the propagation axis.  
Any gravitational wave signal is unchanged by rotating the  binary by $\pi$ around the propagation direction.  
This exact discrete symmetry insures two physically distinct binaries can produce the same gravitational wave signal and
can never be distinguished.  
For this reason, our MCMCs evaluate the polarization angle $\psi_L$ only over  a half-domain $[0,\pi]$: the remaining
half-angle space follows from symmetry.  
Physically, however, at least two physically distinct spin, $\hat{L}$, and $\hat{J}$ configurations produce the same best fitting
gravitational wave signal.    
These two spin configurations can be part of the same probability contour or two discrete islands.

For circularly polarized nonprecessing sources, these two spin configurations blur together: the gravitational wave signal is
independent of $\psi_L-\phi_{orb}$ (for example), preventing independent measurement of $\psi_L$; see, e.g., the top
left panel in Figure 4 of \abbrvAlignedPE{}.  
For generic nonprecessing sources, however, higher harmonics generally isolate the best-fitting gravitational wave
signal and thus polarization angles $\psi_L$, producing two distinct and exactly degenerate islands of probability over $\psi_L\in[0,2\pi]$; see \abbrvAlignedPE{}.
For generic precessing sources, the  same degeneracy applied to the \emph{total} angular momentum produces two distinct, exactly degenerate choices for the
direction $\psi_J$ of $\hat{J}$ in the plane of the sky.   
This degeneracy cannot be broken.  However, at leading order, precessing binaries can still be degenerate in
$\psi_L\pm \phi_{\rm orb}$, in the absence of higher harmonics.  As we will see below, this degeneracy can be broken,
ruling out discrete choices for $\psi_L$.

\section{Parameter estimation of precessing binaries}
\label{sec:Results}

\begin{table*}
\begin{tabular}{lll|ll|lr|rr}
Source & Harmonics & Seed & $\rho$ & $\rho_{\rm rec}$ & $\ln Z$ & $\ln V/V_{\rm prior}$ & $N_{\rm eff}$\\ \hline
{A} & {no} & - & 19.86 & 20.13 & 165. & -37.9 & 10037 \\
 {A} & {no} & 1234 & 19.86 & 21.26 & 186. & -40.1 & 10042 \\
 {A} & {no} & 56789 & 19.86 & 20.59 & 176. & -36.4 & 10110 \\
 {C} & {no} & {-*} & 19.12 & 19.39 & 146. & -41.6 & 101600 \\
 {C} & {no} & {1234*} & 19.12 & 20.64 & 169. & -43.9 & 105941 \\
 {C} & {no} & {56789*} & 19.12 & 18.83 & 138. & -37.9 & 105042 \\
 {C} & {with} & {1234*} & 19.73 & 21.05 & 144. & -41.8 & 42701 \\
 {C} & {with} & {56789*} & 19.73 & 19.32 & 142. & -41.9 & 9814 \\
\end{tabular}

\caption{\label{tab:Simulations}\textbf{Simulations used in this work}:Table of distinct simulations performed.  The
  first set of columns indicate the simulated binary,
  whether higher harmonics were included, and random seed choice used to generate noise 
  (a ``-'' means no noise was used; the asterisk indicates a different noise and MCMC realization).   
The two quantities $\rho_{\rm inj},\rho_{\rm rec}$ provide the
  injected and best-fit total signal amplitude in the network [Eqs. (19) and (22) in \abbrvAlignedPE].
The latter quantity depends on the noise realization of the network. 
The columns for $\ln Z$ and $V/V_{\rm prior}$ provide the evidence [Eq. (15) in \abbrvAlignedPE] and volume fraction
[Eq. (17) of \abbrvAlignedPE]; the evidence, volume fraction, and signal amplitude are related by $\rho_{\rm rec}^2/2  =
\ln Z/(V/V_{\rm prior})$.  
}
\end{table*}

To construct synthetic data containing a signal, to interpret that signal, and to compare interpretations from different
simulations to each other and to theory, we adopt the same methods as used in  \abbrvAlignedPE.  
Specifically, to determine the shape of each posterior, we employ the  \texttt{lalsimulation} and
 \texttt{lalinference} \cite{LIGO-CBC-S6-PE,gw-astro-PE-Raymond} code libraries developed by the LIGO Scientific
 Collaboration and Virgo collaboration.  
As in \abbrvAlignedPE{}, we adopt a fiducial 3-detector network: initial LIGO and Virgo, with analytic gaussian noise
power spectrum provided by their Eqs. (1-2).    
In contrast to the simplified, purely single-spin discussion adopted in Section \ref{sec:Review} to describe the
kinematics of the  \emph{physical signal in the data},  the model used to \emph{interpret} the data allows for  nonzero,
generic spin on \emph{both} compact objects.   %
That said, because compact object spin scales as the mass squared times the dimensionless spin parameter ($S=m^2 \chi$), in our high-mass-ratio systems the small neutron star's spin has
minimal dynamical impact.  
Our simulations  show gravitational waves  provide almost no information about the neutron star's spin
magnitude or direction.  
For the purposes of simplicity, we will omit further mention of the smaller spin henceforth.  

\begin{table*}
\begin{tabular}{ccc|cc|ccc|cccc|ccc|c}
Source & Harmonics & Seed & $ \rho$ & $\hat{\rho}$ & $\sigma_{\mc} $ & $\sigma_{\eta}$ &
$\sigma_{\chi_1}$ & $\sigma_t $ & $\sigma_{RA}$ & $\sigma_{DEC}$ & A & $\sigma_{\alpha_{JL}}$ & $\sigma_{\theta_{JN}}$ & $\sigma_{\beta_{JL}}$ & $N_{\rm eff}$ \\ 
   &  & &%
   & &%
  $\times 10^3$ & $\times 10^3$ &  &%
   ms & deg & deg & $\rm deg^2$ &  & & &  \\
\hline
 {A} & {no} & - & 19.86 & 20.13 & 5.16 & 5.62 & 0.040 & 0.505 & 0.493 & 0.747 & 1.06 &     0.094 & 0.0880 & 0.0594 & 10037 \\
 {A} & {no} & 1234 & 19.86 & 21.26 & 5.06 & 4.33 & 0.041 & 0.491 & 0.542 & 0.733 & 1.21 & 0.100 & 0.0891 &0.0643 & 10042 \\
 {A} & {no} & 56789 & 19.86 & 20.59 & 5.08 & 4.31 & 0.033 & 0.313 & 0.572 & 0.714 & 1.27 & 0.105 &0.0807 & 0.0523 &  10110 \\
 {C} & {no} & {-*} & 19.12 & 19.39 & 4.81 & 4.39 & 0.032 & 0.276 & 0.464 & 0.739 & 0.967 & 0.105 &0.0741 & 0.0484 & 101600 \\
 {C} & {no} & {1234*} & 19.12 & 20.64 & 4.76 & 3.72 & 0.032 & 0.247 & 0.385 & 0.647 & 0.709 & 0.0937 & 0.0624 & 0.0473 & 105941 \\
 {C} & {no} & {56789*} & 19.12 & 18.83 & 5.40 & 4.23 & 0.039 & 0.221 & 0.469 & 0.717 & 0.960 & 0.113 & 0.0784 & 0.0622 & 105042 \\
 {C} & {with} & {1234*} & 19.73 & 21.05 & 4.80 & 3.49 & 0.030 & 0.191 & 0.309 & 0.551 & 0.474 & 0.087 & 0.058 &0.045 & 42701 \\
 {C} & {with} & {56789*} & 19.73 & 19.32 & 4.87 & 4.20 & 0.039 & 0.191 & 0.393 & 0.661 & 0.651 & 0.112 & 0.0752 & 0.0603 & 9814
\end{tabular}

\caption{\label{tab:UncertaintyReport}\textbf{One-dimensional parameter errors}: Measurement accuracy $\sigma_x$ for
$x$ one of   several intrinsic ($\mc,\eta, \chi_1$), extrinsic ($\psi_{\pm},t,RA,DEC$), and precession-geometry ($\alpha_{JL},\beta_{JL},\theta_{JN}$) parameters.    
The extrinsic parameters are  the event time
  $t$; the sky position measured in RA and DEC; and the sky area $A$, estimated using the $2\times 2$ covariance matrix
  $\Sigma_{ab}$ on the sky via $\pi |\Sigma|$.  %
The precession cone parameters are as described in Figure \ref{fig:Coords}: the precession phase $\alpha_{JL}$ at the reference frequency; the precession cone
opening angle $\beta_{JL}$; and the viewing angle $\theta_{JN}$.  
}
\end{table*}

\subsection{Intrinsic parameters}
\begin{figure}
\includegraphics[width=\columnwidth]{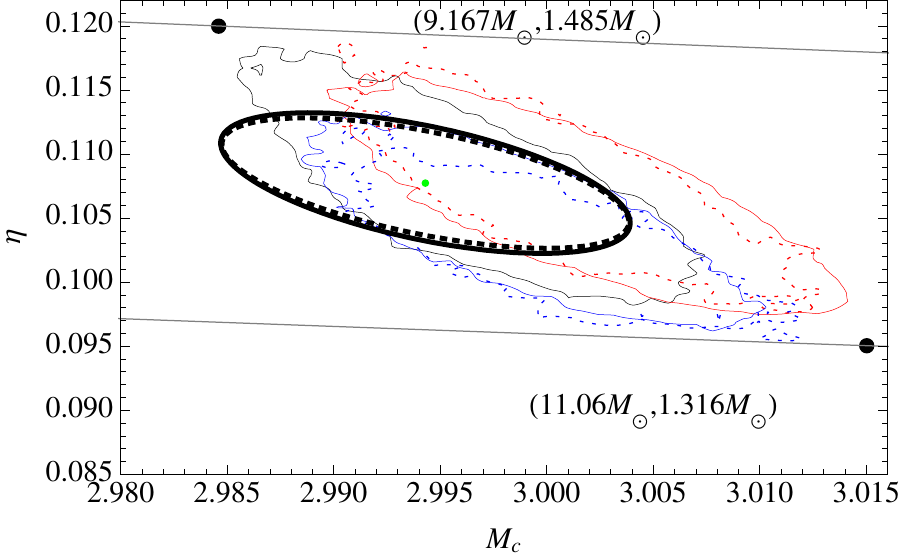}
\includegraphics[width=\columnwidth]{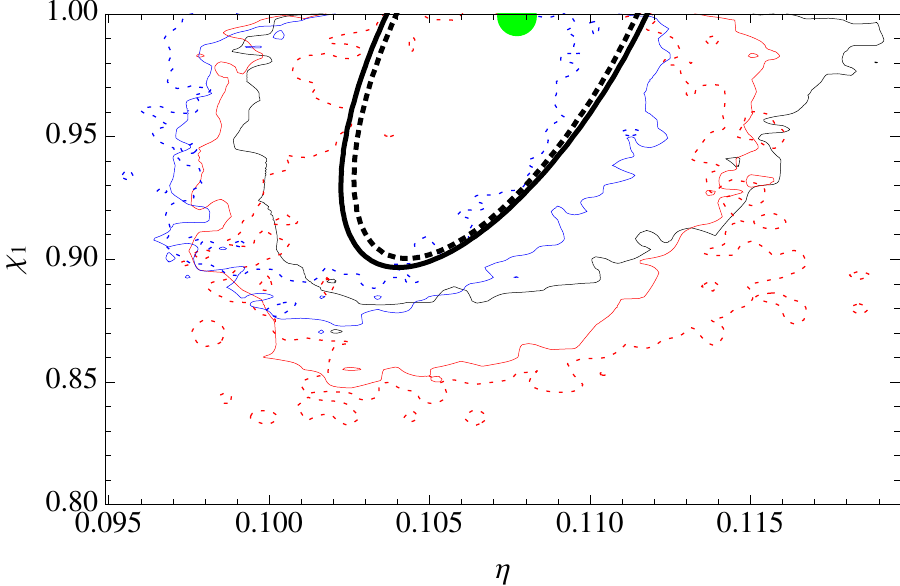}
\includegraphics[width=\columnwidth]{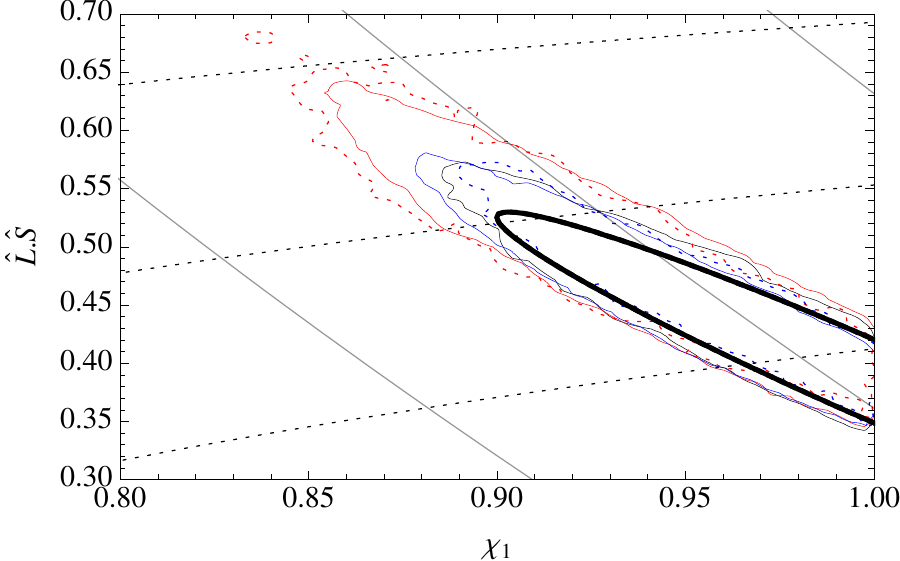}
\caption{\label{fig:Results:Intrinsic}\textbf{Estimating astrophysical parameters (C)}:   For our
  fiducial binary C, the solid and dotted lines show an estimated 90\% confidence interval with and without higher harmonics, respectively; colors indicate different noise
realizations; and the (nearly indistinguishable) thick solid and dashed lines shows an approximate  effective Fisher
matrix result, with and without higher harmonics, not accounting for the constraint imposed by $\chi_1<1$.   Results for case A are qualitatively and quantitatively similar.  
The different panels show different two-dimensional projections of the astrophysically relevant parameters of a merging BH-NS
  binary: the binary mass ratio, black hole spin, and degree of spin-orbit misalignment $\kappa \equiv \hat{L}\cdot \hat{S}_1$. 
\emph{Top,center panels}:  The masses and spin magnitude  of the binary can be measured very reliably,
consistent with a single gaussian distribution in four dimensions.   The analytic predictions produced by an effective
Fisher matrix agree qualitatively but not quantitatively with our simulations.
\emph{Bottom panel}: To guide the eye, the posterior versus  $\chi_1$ and $\hat{L}\cdot \hat{S}_1$ is compared with
contours of constant $\beta_{JL}=\cos^{-1}0.65, 0.7, 0.75$ (precession cone opening angle; dotted black) and $\Omega_p$ [Eq. (\ref{eq:OmegaPrec})]
(precession rate; solid black).    %
The precession rate is relatively well
constrained by the presence of several ($\simeq \NprecSpecific$) precession cycles available in data, while the geometry is relatively poorly constrained,
relative to the whole $\chi_1$ vs $\hat{L}\cdot \hat{S}_1$ plane.
}
\end{figure}

As shown  in Figure \ref{fig:Results:Intrinsic}, the intrinsic parameters of our relatively loud ($\rho \simeq 20$)  fiducial binaries are extremely
well-constrained.  
For example, the neutron star's mass, black hole's mass, and black hole spin are all relatively well-measured, compared to the accuracy of
existing measurements and hypothesized distributions of these parameters  \cite{obs-ns-masses-Ozel2012,2012ApJ...757...36K,2013ApJ...766L..14H}.  
Higher harmonics provide relatively little additional information about these parameters.

Applied to an even simpler idealized problem -- a similar source known to be directly overhead two orthogonal detectors
-- the effective fisher matrix procedure of \abbrvEFM{}  produces qualitatively similar results, notably reproducing  relatively minimal impact from
higher harmonics.
Given the simplifications adopted,  the effective fisher matrix predictions inevitably disagree quantitatively with our detailed Monte Carlo
calculations, particularly regarding multidimensional correlations.  We nonetheless expect the effective fisher matrix to
correctly identify  \emph{scales} and \emph{trends} in parameter estimation; moreover, being amenable to analysis, this
simple construct allows us to develop and validate simple interpretations for why some parameters can be measured as
well as they are.     As a concrete example, we can explain Figure \ref{fig:Results:Intrinsic}.
The time-dependent orbital phase depends on the black hole spin, principally through the ``aligned component'' $\hat{L}\cdot
\vec{S}_1$ \cite{1995PhRvD..52..848P}.   As discussed in \abbrvEFM{} and
\abbrvAlignedPE{}, the ``aligned component'' cannot be easily distinguished from the
mass ratio in the absence of precession.  Spin-orbit precession breaks this degeneracy, allowing significantly tighter
constraints on the mass ratio of our precessing binary.  In our particular example, comparing  Fig 3 in \abbrvAlignedPE{} to our Figure
\ref{fig:Results:Intrinsic},  we can measure $\eta$ and hence the smaller mass roughly three times more
accurately at the same signal amplitude.  
As seen in the bottom panel of Figure \ref{fig:Results:Intrinsic}, both the spin-orbit
misalignment $\hat{L}\cdot\hat{S}_1$ and spin magnitude $\chi_1$ remain individually poorly-constrained.  As a concrete
example, our ability to measure $\chi_1$ for this precessing binary is comparable to the accuracy possible for a similar
nonprecessing binary [\abbrvAlignedPE].   

One correlated combination of $\hat{L}\cdot \hat{S}_1$ and $\chi_1$ is well-constrained: the combination that enters into
the precession rate.  In Figure \ref{fig:Results:Intrinsic} we show contours of constant precession cone opening angle
($\beta_{JL}$) and constant precession rate [LO Eq. (7-8)]
\begin{eqnarray}
\label{eq:OmegaPrec}
\Omega_p = \frac{|J|}{2r^3} =\eta \left(2+\frac{3 m_2}{2m_1} \right)v^5\sqrt{1+2\kappa\gamma+\gamma^2}
\end{eqnarray} %
When evaluating these
expressions, we estimate $\gamma\simeq 1.85 \chi_1$ [Eq.   (\ref{eq:def:gamma})], so the contours shown correspond   to
$\cos \beta_{JL} =0.65,0.7,0.75$ and $\sqrt{1+2\kappa\gamma+\gamma^2} = 2.2, 2.4,2.6$.  
As expected, the presence of several precession cycles allows us to relatively tightly constrain the precession rate.
Future gravitational wave detectors, being sensitive to longer signals and hence more precession cycles, can be expected
to even more tightly constrain this combination.   By contrast,
as described below,  the precession geometry $\beta_{JL}$ is relatively poorly constrained, with error
independent of the number of orbital or precession cycles.\footnote{%
With relatively few precession cycles in our study, the discrepancy between these two measurement accuracies is fairly
small.  However, when advanced instruments with longer waveforms can probe more precession cycles, we expect this simple
argument will explain dominant correlations.}

\ForInternalReference{
\begin{figure}
\includegraphics[width=\columnwidth]{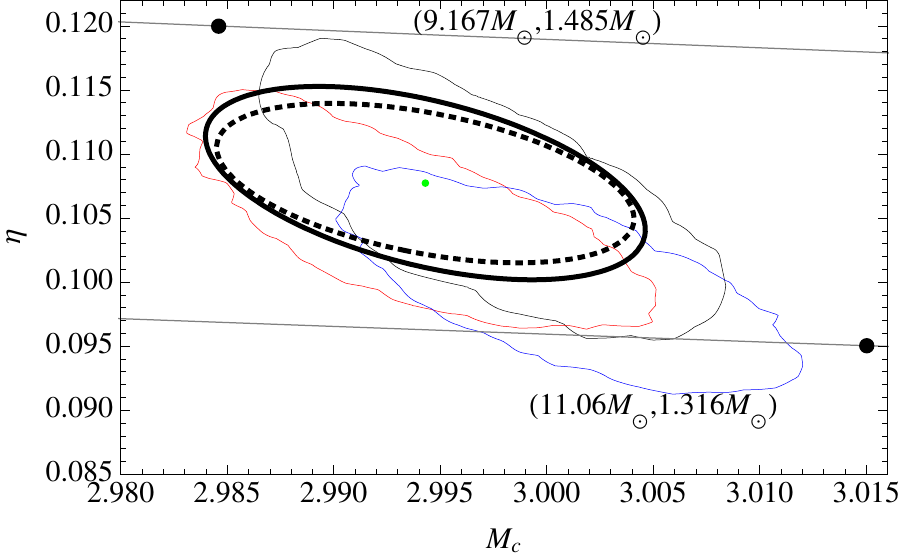}
\includegraphics[width=\columnwidth]{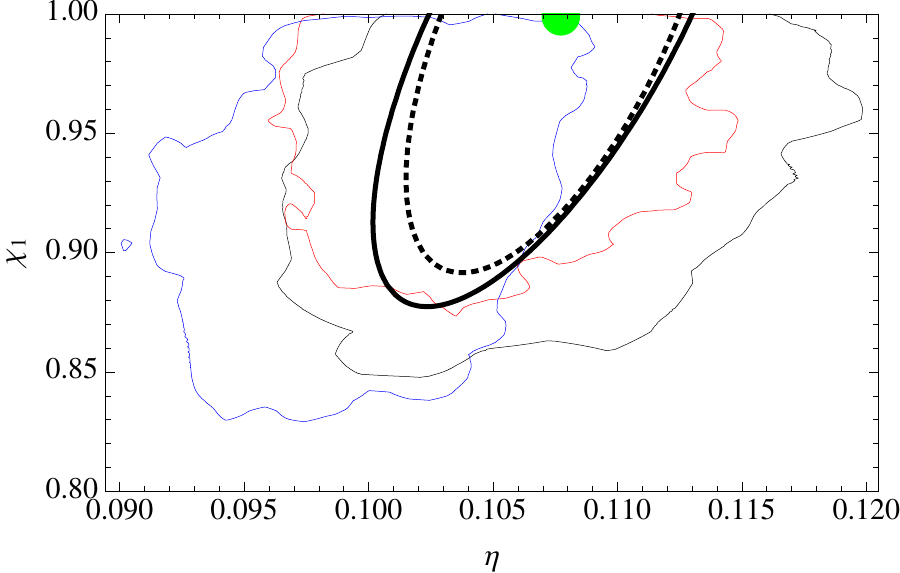}
\includegraphics[width=\columnwidth]{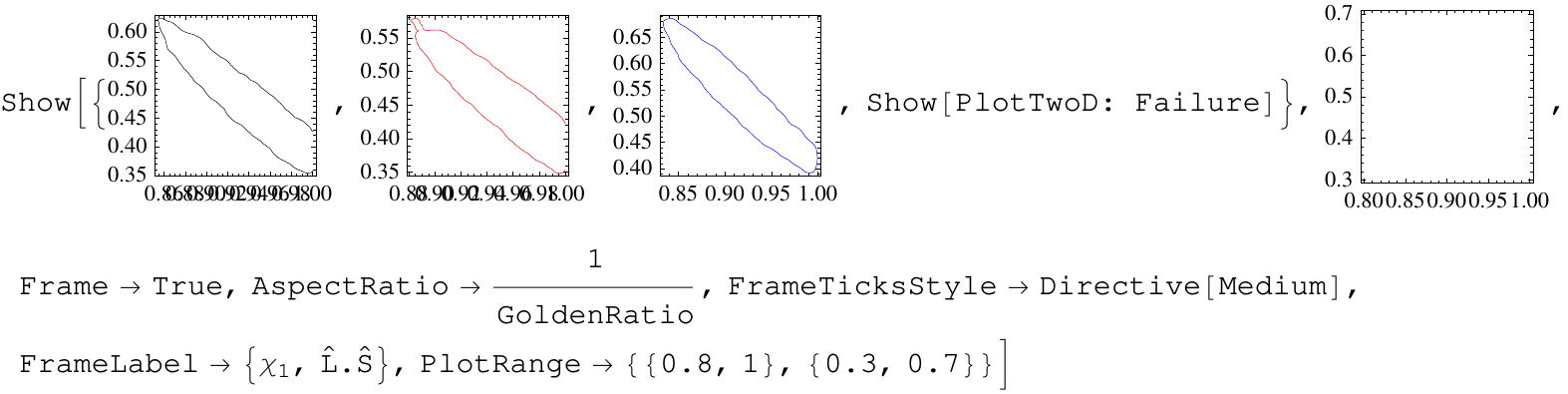}
\caption{\label{fig:Results:Intrinsic:A}\textbf{Estimating astrophysical parameters (A)}:
As Figure \ref{fig:Results:Intrinsic}, except  the effective Fisher matrix predictions and Monte Carlo results for case
A are shown.    \editremark{Unlike case C}, higher harmonics seem to significantly improve our ability to estimate
intrinsic parameters; see also Table \ref{tab:UncertaintyReport}.  \editremark{Unfortunately, case A corresponds to
  apples and oranges: the physical systems are different with and without higher harmonics}
}
\end{figure}
}

\subsection{Geometry}
\begin{figure*}
\includegraphics[width=\columnwidth]{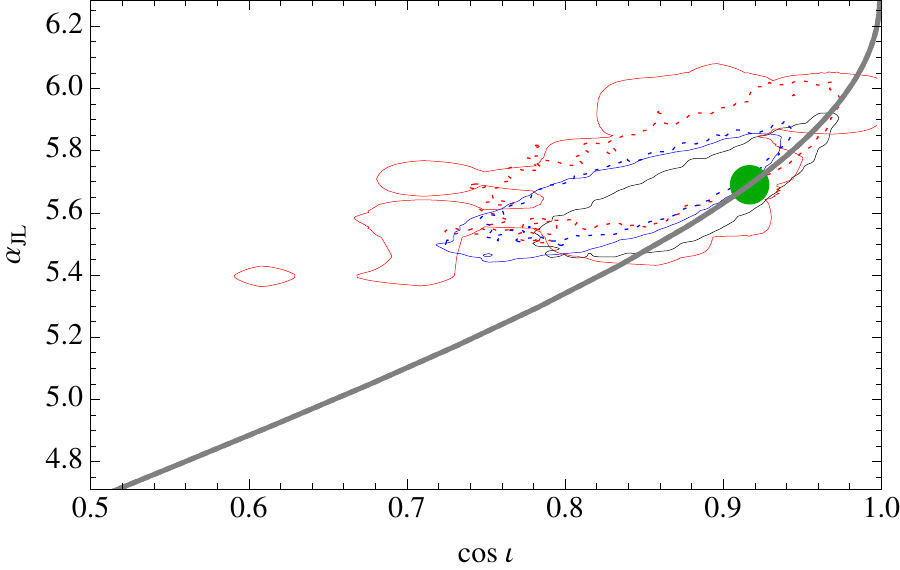}
\includegraphics[width=\columnwidth]{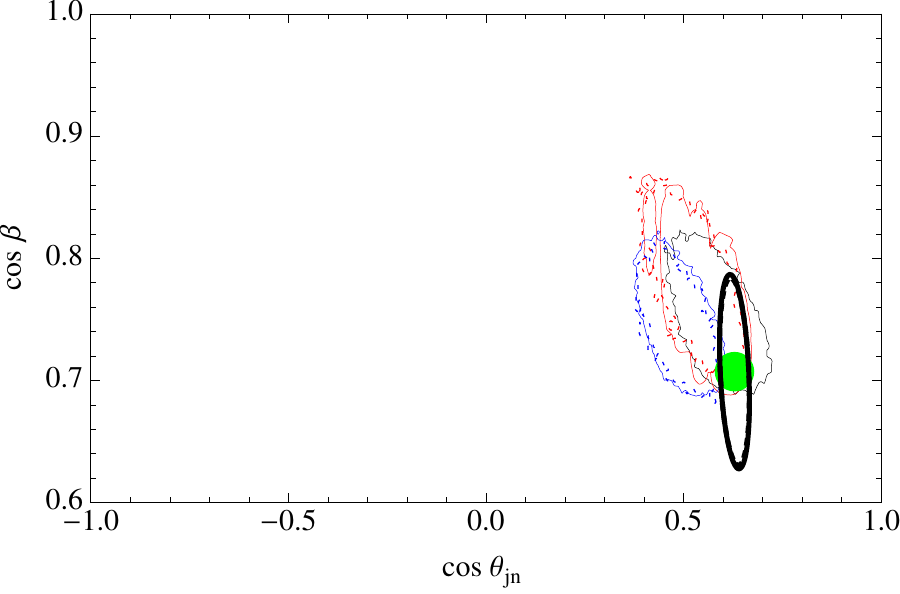}
\includegraphics[width=\columnwidth]{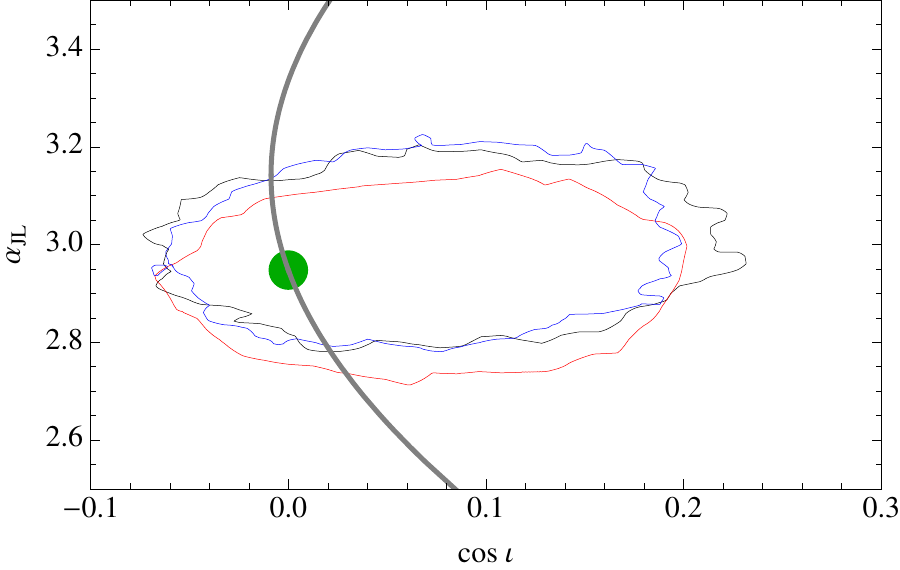}
\includegraphics[width=\columnwidth]{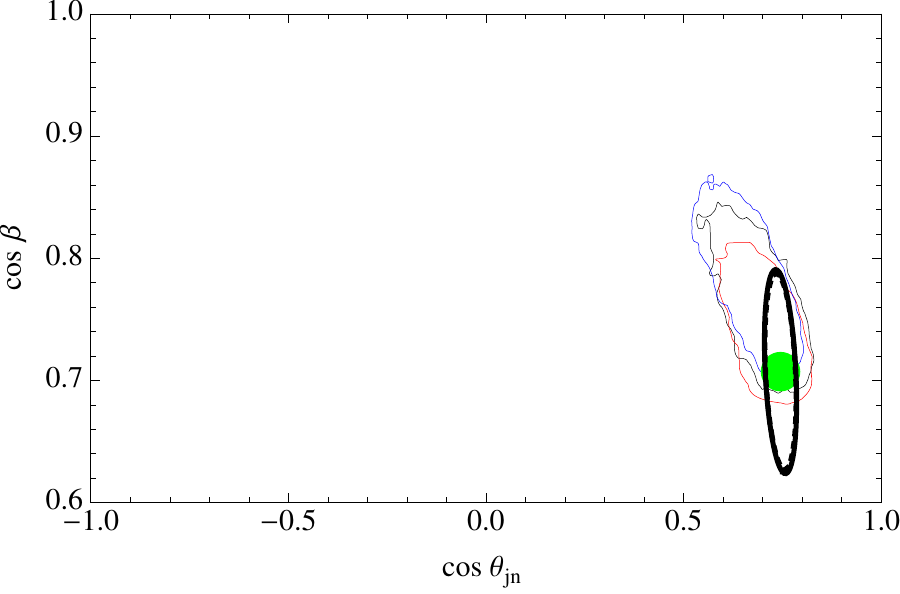}
\caption{\label{fig:Geometry:AngularMomenta}\textbf{Source geometry: Angular momenta (C,A)}: 
For case C (top panels) and case A (bottom panels), the posterior for the ``precession cone'' (path of the angular momentum direction), expressed using the
precession cone representation.   This figure demonstrates that both the path ($\theta_{JN},\beta_{JL}$) and instantaneous
orientation ($\alpha_{JL},\iota$)  of the orbital angular momentum can be well-determined 
As in Figure \ref{fig:Results:Intrinsic}, colors indicate different noise realizations; solid and dotted lines indicate
the neglect or use of higher harmonics; the green point shows the actual value; and the solid gray path shows the
trajectory of $L$  over one precession cycle.  
\emph{Left panels}:  The precession angle $\alpha_{JL}$ of $L$ around $J$.  For comparison, the green
  points show the simulated values; when present, the solid blue path shows variables covered in one precession cycle.   
Roughly speaking, the precession phase can be measured with relative accuracy  tens of
percent at this signal amplitude $\rho$.  
\emph{Right panels}: Illustration that both the opening angle $\beta_{JL}$ of the precession cone and the angle
$\theta_{JN}$ between the line of sight and $\hat{J}$ can be measured accurately.
}
\end{figure*}

As expected analytically and demonstrated by Figure \ref{fig:Geometry:AngularMomenta}, precession-induced modulations
encode the orientation of the various angular momenta relative to the line of sight.  
For our loud fiducial signal, the individual spin components can be well-constrained.  
Equivalently, because our fiducial source performs many precession cycles about a wide precession cone and because that
source is viewed along a generic line of sight, we can tightly constrain the precession cone's geometry: its opening
angle; its orientation relative to the line of sight; and even the precise precession phase, measured either by $\cos
\iota$ or $\alpha_{JL}$.  
The effective Fisher matrix provides a reliable estimate of how well these parameters can be measured; see Table
\ref{tab:UncertaintyReport} and Figure
\ref{fig:Geometry:AngularMomenta}.   

\subsection{Comparison to and interpretation of analytic predictions}

\abbrvEFM{} presented an effective Fisher matrix for two fiducial precessing binaries, adopting a specific
post-Newtonian model to evolve the orbit.  Following \abbrvAlignedPE{}, we adopt a refined post-Newtonian model,
including higher-order spin terms.  In the Supplementary Material, available online, we provide a revised effective
Fisher matrix, including the contribution from these terms.   
Table \ref{tab:RevisedFisher:A} summarizes key features of this seven-dimensional effective Fisher matrix for case A.
As noted above, the two-dimensional marginalized predictions are in good qualitative agreement.  The one-dimensional marginalized predictions agree surprisingly well with our simulations [Table
  \ref{tab:UncertaintyReport}].  
Since the ingredients of the effective Fisher matrix are fully under our analytic control, we can directly assess what
factors drive measurement accuracy in each parameter.

\begin{table}
\begin{tabular}{l|l}
Property & Value(s) \\\hline
$\lambda_k$   & 6412, 673, 87, 5.5,0.76, 0.27, 0.004 \\\hline
$\sigma_{\mc}$ & 0.0048 $M_\odot$ \\
$\sigma_{\eta}$ &  0.0035 \\
$\sigma_{\chi_1}$ &  0.057 \\
$\sigma_{\beta_{JL}}$ & 0.06 \\
$\sigma_{\alpha_{JL}}$ & 0.10\\
$\sigma_{\theta_{JN}}$ & 0.07 \\
$\sigma_{\psi_J}$ ($\sigma_{\phi_{\rm ref}}$ ) & 0.09 (0.78) \\\hline
\end{tabular}
\caption{\label{tab:RevisedFisher:A}\textbf{Properties of precessing effective Fisher matrix}: Quantities derived from the normalized effective
  Fisher matrix $\hat{\Gamma}$, as provided in the supplementary information:  the eigenvalues   $\lambda_k$ and
  one-dimensional parameter measurement accuracies $\sqrt{\hat{\Gamma}^{-1}/\rho^2}$ evaluated for $\rho=20$. 
  (As we only compute Fisher matrices after marginalizing over $\psi$ or $\phi_{\rm ref}$, we provide only 7 eigenvalues
  and independent parameter measurement errors at a time.)
}
\end{table}

First and foremost, as in \abbrvEFM{}, this effective Fisher  matrix has a hierarchy of scales and eigenvalues, with decreasing measurement
error: $\mc,\eta,\chi, \ldots$.  Unlike nonprecessing binaries, this hierarchy does not clearly split between
 well-constrained intrinsic parameters ($\mc,\eta,\chi_1$) and poorly-constrained geometric parameters (everything else); for
 example, as
 seen in Table \ref{tab:RevisedFisher:A}, the eigenvalues of the Fisher matrix span a continuous range of scales. 

The scales in the Fisher matrix are intimately tied to timescales and angular scales in the outgoing signal.  The
largest eigenvalues  of the Fisher matrix are  set by the shortest timescales: the orbital timescale, and changes to the orbital phase versus
time.  These scales control measurement of $\mc,\eta, L\cdot a$ and set the reference event time and phase.  Qualitatively speaking, we measure these parameters
well because good matches require the orbital phase to be aligned over a wide range in time.    We measure the reference
waveform phase reliably because each waveform must be properly aligned.  For this reason, parameters related to orbital
phase (i.e., $\mc$) can be measured to order $1/\sqrt{N_{cycles}}$, times suitable powers of $v$ to account for the
post-Newtonian order at which those terms influence the orbital phase.

 The next-shortest scales are precession scales: changes to the zero of precession phase, and how precession phase accumulates with time.  Qualitatively
 speaking, we can measure the reference precession phase reliably because each \emph{precession cycle} needs to be in
 phase.  %
Due to spin-orbit precession, our fiducial  BH-NS binaries will undergo   $N_{\rm cycles}\simeq 10 $ ($30$) amplitude and
phase modulations in band, as seen by an  initial (advanced) detector [\abbrvBLO{} Eq. (9), for an
  angular-momentum-dominated binary, with $|L|>|S|$]
\begin{eqnarray}
\label{eq:PrecessionCycles}
N_P &\simeq&  \int_{\pi f_{min}}^{\pi f_{max}} d f_{orb} \frac{dt}{d f_{orb}} \Omega_p \nonumber \\
 &=&  \frac{5}{96}(2+1.5 \frac{m_2}{m_1}) [(M\pi f_{min})^{-1} - (M\pi f_{max})^{-1}] \nonumber \\
&\approx & \frac{10 (1+0.75 m_2/m_1) }{M/10 M_\odot} (f_{\rm min}/50\unit{Hz})^{-1}
\end{eqnarray}
This estimate agrees favorably with the roughly \NprecSpecific{} precession cycles performed by our spin-dominated ($|S|>|L|$)
binary between 30 and 500 Hz [Figure
  \ref{fig:AnglesVersusTime}].  
For this reason, parameters tied to spin-orbit precession \emph{rate} $\Omega_p$ (i.e., $\eta$) will be measured to a relative
accuracy $1/\sqrt{N_p}$.  Applied to the mass ratio, this estimate leads to the surprisingly successful estimate
\begin{eqnarray}
\label{eq:PrecessionTiedQuantities}
\sigma_{\eta} \simeq O(1)\times \frac{\eta}{\sqrt{N_p}\rho} \simeq O(1)\times 1.6\times 10^{-3}
\end{eqnarray}
i.e.,  roughly $1/\sqrt{N_P}$ times smaller than the measurement accuracy possible without breaking the spin-mass
ratio degeneracy.

While some parameters change the rate at which orbital and precession phase accumulate, other reference phases simply
fix the geometry.   For example,  a shift in the precession phase at some reference frequency (i.e., $\alpha(f=100\unit{Hz})$)  leads to a
\emph{correlated} shift in the precession and hence gravitational wave phase in each precession cycle.  In other words,
like our ability to measure the orbital phase at some time, our ability to measure the reference precession phase is
essentially independent of the number of orbital or precession cycles, solely reflecting geometric factors.
We expect the accuracy with which these purely geometric parameters  $x$ can be determined can be estimated from
first principles.    To order of magnitude, we expect  Fisher matrix 
components $\Gamma_{xx}$ comparable to $\Delta x^2$, where $\Delta x$ is %
the parameter's range. 
 For
example, the angular parameters ($\beta_{JL},\theta_{JN},\alpha,\phi$) should be measured to within 
\begin{eqnarray}
\sigma_{\rm angle} \simeq \frac{(2\pi)}{\sqrt{12} \rho} \simeq 0.09 \unit{rad} 
\end{eqnarray}    
where the factor $\sqrt{12}$ is the standard deviation of a uniform distribution over $[0,1]$.  
This simple order-of-magnitude estimate compares favorably to the Fisher matrix results shown in Table \ref{tab:RevisedFisher:A} and to our full
numerical simulations [Figure \ref{fig:Geometry:AngularMomenta} and Table \ref{tab:UncertaintyReport}].   
This naive estimate ignores all dependence on precession geometry; in general, all geometric factors are tied directly to the magnitude of precession-induced modulations, which grow increasingly significant for larger misalignment, roughly in proportion
to  $\cos \beta_{JL}$.   
This estimate for how well geometric angles can be measured should break down for nearly end-over-end precession ($\beta_{JL} \rightarrow \pi/2$).  Nearly
end-over-end precession requires extreme fine tuning; is associated with transitional precession;  and is correlated
with  rapid change in $\beta_{JL}$ [\abbrvBLO].  We anticipate a different set of approximations will be required
to address this limit.

\subsection{Relative role of higher harmonics }
\begin{figure*}
\includegraphics[width=0.3\textwidth]{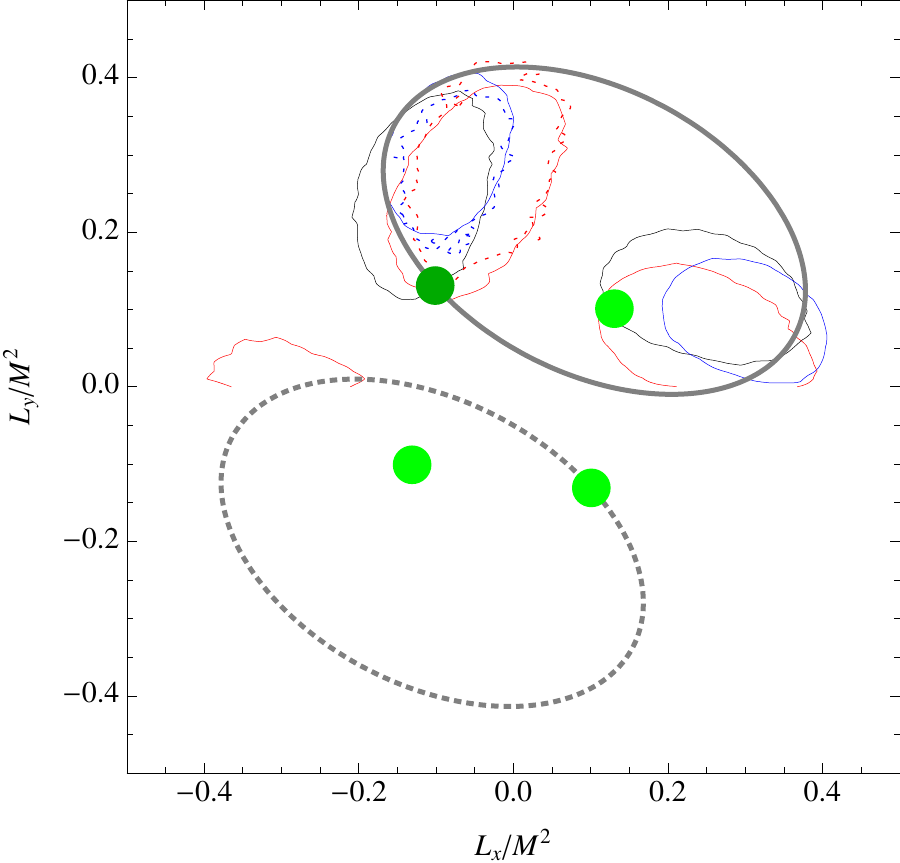}
\includegraphics[width=0.3\textwidth]{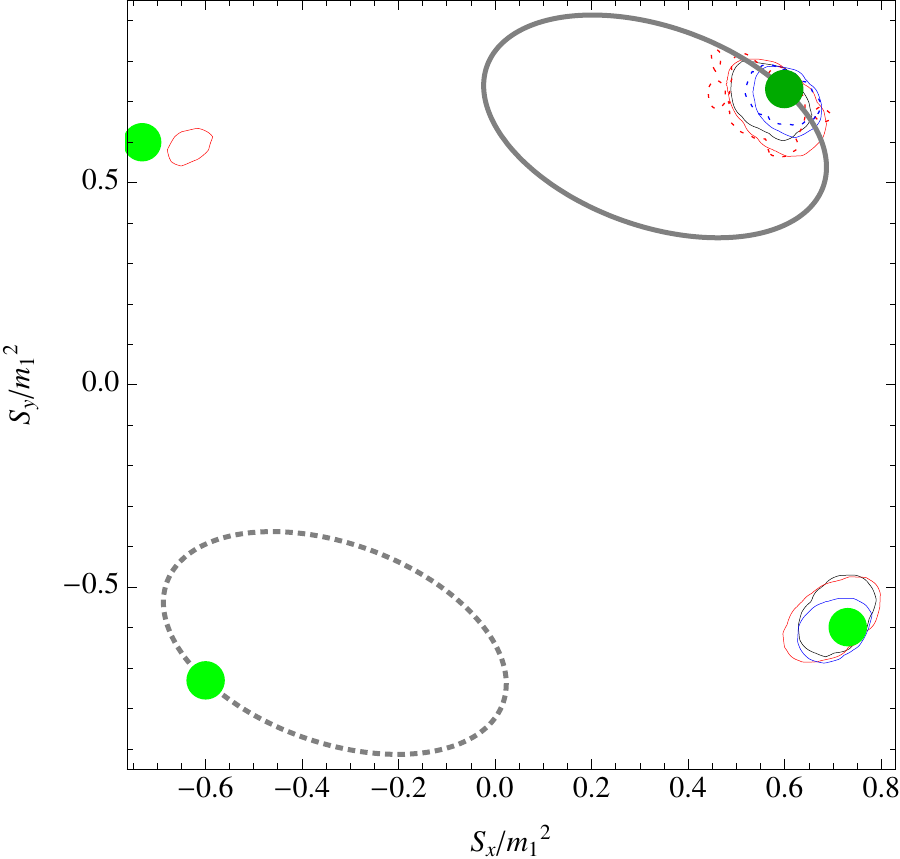}
\includegraphics[width=0.3\textwidth]{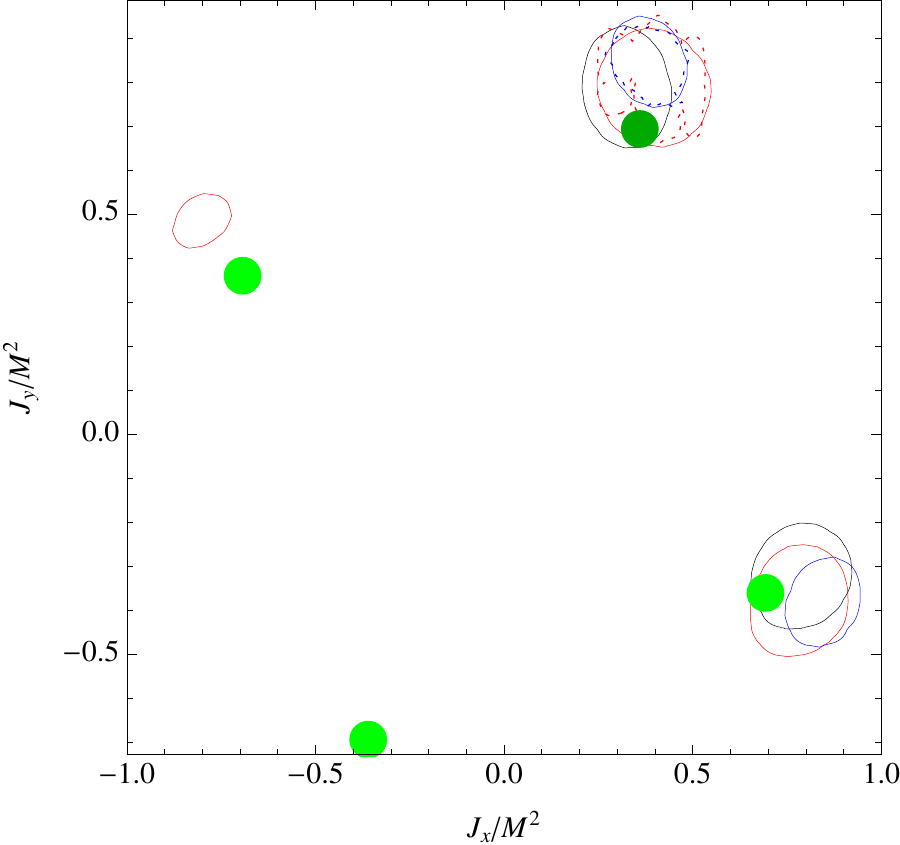}
\caption{\label{fig:AngularMomentumOnSkyPosterior}\textbf{Angular momentum direction on the sky (C) }: Projection of the orbital angular momentum direction
  ($\hat{L}$) on the plane of the sky at $f=100\unit{Hz}$; compare to Figure \ref{fig:AnglesVersusTime}.
This figure demonstrates that  the individual angular momenta to be
well-constrained to \textbf{two} discrete regions and that higher harmonics allow us to distinguish between the two
alternatives; and (3) that the precession cone is well-determined, at the accuracy level expected from the number of
precession cycles.   
As in Figure \ref{fig:Results:Intrinsic}, colors indicate different noise realizations; solid and dotted lines indicate
the neglect or use of higher harmonics; and the green point shows the expected solution.
}
\end{figure*}

To this point, both our analytic and numerical calculations suggest higher harmonics provide relatively little
additional information about intrinsic and extrinsic parameters.   
That said, as illustrated by Figure \ref{fig:AngularMomentumOnSkyPosterior}, higher harmonics do break a discrete 
 degeneracy, determining the   orientation of $\hat{L}$ on the plane of the sky at $f=100\unit{Hz}$ up to a rotation by $\pi$.  
\abbrvAlignedPE{} used the evidence to demonstrate conclusively that higher harmonics had no additional impact, beyond
improving knowledge of one parameter.   
Given expected systematic uncertainties in the evidence, at the present time we do not feel we can make as robust and
global a statement.  
That said, all of our one- and two-dimensional marginalized posteriors support the same conclusion: higher harmonics
provide little new information, aside from breaking one global degeneracy.

\subsection{Timing, sky location, and distance}
\begin{figure}
\includegraphics[width=\columnwidth]{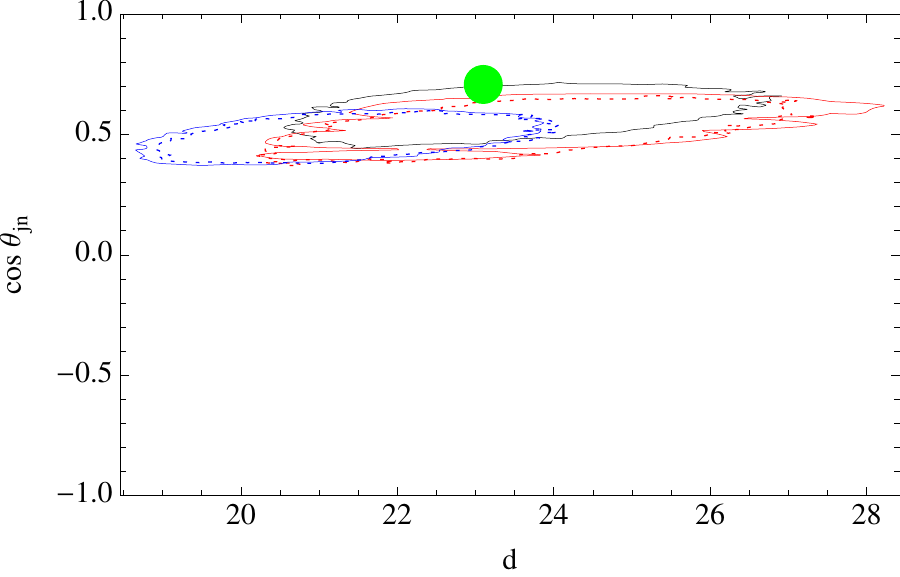}
\caption{\label{fig:DCosIota}\textbf{Distance and inclination degeneracy broken (C)}: Posterior probability contours in distance and
  inclination.
}
\end{figure}

As seen in Figure \ref{fig:DCosIota}, precessing binaries do not have the strong source orientation versus distance degeneracy that plagues nonprecessing
binaries: because they emit distinctively different multi-harmonic signals in each direction, both the distance and emission
direction can be tightly constrained.   

Conversely, the sky location of precessing binaries can be determined to little better than the sky location of a
nonprecessing binary with comparable signal amplitude; compare, for example, Table \ref{tab:UncertaintyReport} and Figures \ref{fig:Geometry:AngularMomenta}
against the corresponding figures in \abbrvAlignedPE.

Finally, the event time can be marginally better determined for a precessing than for a nonprecessing binary.  
This accuracy may be of interest for multimessenger observations of gamma ray bursts.

\subsection{Advanced versus initial instruments}
All the discussion above  assumed first-generation instrumental sensitivity.  For comparison
and to further validate our estimates, we have also done one calculation using the expected sensitivity of second-generation
instruments \cite{LIGO-aLIGODesign-Sensitivity,2010CQGra..27q3001A}.  In this calculation, the source (event C) has been
placed at a larger distance ($d=298.7 \unit{Mpc}$) to produce the same network SNR.
Also  unlike the analysis above, we have for simplicity  assumed the smaller compact object has no spin.  

\begin{figure}
\includegraphics[width=\columnwidth]{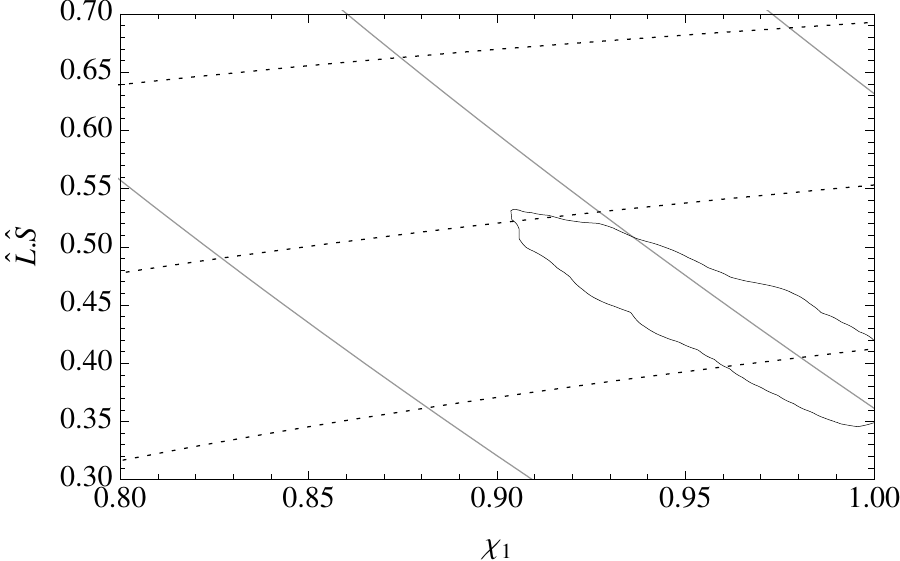}
\caption{\label{fig:Results:Intrinsic:Advanced}\textbf{Estimating astrophysical parameters with advanced detectors}:
  Like the bottom panel of Figure \ref{fig:Results:Intrinsic}, but using advanced instruments; see  Table \ref{tab:CompareInitialAdvanced}.
}
\end{figure}

Table \ref{tab:CompareInitialAdvanced} shows the resulting one-dimensional measurement accuracies, compared against a
concrete simulation.  All results agree with the expected scalings, as described previously.  First and foremost,   all geometric quantities ($RA, DEC,\alpha_{JL},\beta_{JL},
\theta_{JN}$) and time can be measured to the same accuracy as in initial instruments, at fixed SNR.  
Second, quanitites that
influence the orbital decay -- chirp mass, mass ratio, and spin -- are all measured more precisely, because more
gravitational wave cycles contribute to detection with advanced instruments.   
Finally, as illustrated by Figure \ref{fig:Results:Intrinsic:Advanced}, quantities that reflect precession-induced
modulation - the precession rate $\Omega_p$ and precession cone angle $\beta_{JL}$ misalignment -- are at best measured marginally more accurately, reflecting the
relatively small increase in number of observationally-accessible precession cycles for advanced detectors [Eqs. (\ref{eq:PrecessionCycles} and
\ref{eq:PrecessionTiedQuantities}].   
As shown by the bottom panel of Figure \ref{fig:Results:Intrinsic}, this small increase in sensitivity is comparable to
the typical effect of different noise realizations.

\begin{table*}
\begin{tabular}{cccc|cc|ccc|cccc|ccc|c}
Source & Instrument & Harmonics & Noise & $ \rho$ & $\hat{\rho}$ & $\sigma_{\mc} $ & $\sigma_{\eta}$ &
$\sigma_{\chi_1}$ & $\sigma_t $ & $\sigma_{RA}$ & $\sigma_{DEC}$ & A & $\sigma_{\alpha_{JL}}$ & $\sigma_{\theta_{JN}}$ & $\sigma_{\beta_{JL}}$ & $N_{\rm eff}$ \\ 
 &   &  & &%
   & &%
  $\times 10^3$ & $\times 10^3$ &  &%
   ms & deg & deg & $\rm deg^2$ &  & & &  \\
\hline
 {C} & Initial & {no} & no & 19.12 & 19.39 & 4.81 & 4.39 & 0.032 & 0.276 & 0.464 & 0.739 & 0.967 & 0.105 &0.0741 &
 0.0484 & 101600 \\
 {C} & Advanced & {no} & yes & 19.54 & 19.57 & 3.66 & 2.31 & 0.029 & 0.228 & 0.421 & 0.704 & 0.791 & 0.105 &0.0699 & 0.0415 & 4004 \\
\end{tabular}
\caption{\label{tab:CompareInitialAdvanced}\textbf{Parameter estimation with initial and advanced instruments}:
 Like Table \ref{tab:UncertaintyReport},  measurement accuracy $\sigma_x$ for several intrinsic and extrinsic
 parameters.  The first row provides results for initial-scale instruments, duplicating an entry in Table
 \ref{tab:UncertaintyReport}.  The second row provides results for advanced detectors, operating at design sensitivity.  
At fixed signal amplitude, most geometric quantities can be measured to fixed accuracy, independent of detector
sensitivity.  Quantities impacting the orbital phase versus time (mass, mass ratio, and spin) are more accurately
measured with advanced instruments, with their access to lower frequencies and hence more cycles.    
}
\end{table*}

\section{Conclusions}
\label{sec:Conclusions}

In this work we performed detailed parameter estimation for two selected BH-NS binaries, explained several features in
terms of the binary's kinematics and geometry, and compared our results against analytic predictions using the methods
of \cite{gwastro-mergers-HeeSuk-FisherMatrixWithAmplitudeCorrections,gwastro-mergers-HeeSuk-CompareToPE-Aligned}.
First, despite adopting a relatively low-sensitivity initial-detector network for consistency with prior work, we find by example
that parameter estimation of precessing binaries can draw astrophysically interesting conclusions.  Since our study
adopted  relatively band-limited initial detector noise spectra, we expect advanced interferometers \cite{LIGO-Inspiral-Rates,2010CQGra..27h4006H} will perform at
least as well (if not better) at fixed SNR.   For our fiducial
binaries, the mass parameters are constrained well enough to definitively say if it is a BH-NS
binary (as opposed to BH-BH); the mass parameters are constrained better than similar non-precessing binaries; and
several parameters related to the spin and orientation of the binary can be measured with reasonable accuracy. 
Second and more importantly, we were able to explain our results qualitatively and often quantitatively using far
simpler, often analytic calculations. Building on prior work by \abbrvBLO, \abbrvLO, and others
\cite{gwastro-mergers-nr-Alignment-ROS-Polarization}, we argued precession introduced distinctive amplitude, phase, and
polarization modulations on a precession timescale, effectively providing another information channel independent from
the usual inspiral-scale channel found in non-precessing binaries. Though our study targeted only two specific
configurations, we anticipate many of our arguments explaining the measurement accuracy of various parameters can be
extrapolated to other binary configurations and advanced detectors. The effective Fisher matrix approach  of \abbrvEFM{} and
\abbrvAlignedPE{} provides a computationally-efficient means to undertake such extrapolations. 
Third and finally, we demonstrated that for this mass range and orientation, higher harmonics have minimal local but
significant global impact. For our systems, we found higher harmonics broke a degeneracy in the orientation of $\hat{L}$
at our reference frequency (100 Hz), but otherwise had negligible impact on the estimation of any other parameters.

Due to the relatively limited calculations of spin effects in post-Newtonian theory, all inferences regarding black hole
spin necessarily come with significant systematic limitations.  For example,
\citet{gwastro-mergers-pn-spinterms-Nitz2013} imply that poorly-constrained  spin-dependent contributions to the
orbital phase versus time could significantly impact parameter estimation of nonprecessing black hole-neutron star binaries.
Fortunately, the leading-order precession equations and physics are relatively well-determined.  For example, the amplitude of
precession-induced modulations is set by the relative magnitude and misalignment of $\vec{L}$ and $\vec{S}_1$.   In our
opinion, the leading-order symmetry-breaking  effects of precession are less likely to be susceptible to systematic error than
high-order corrections to the orbital phase.   Significantly more study would be needed to validate this hypothesis. 

Robust though these correlations may be, the quantities gravitational wave measurements naturally provide (chirp mass;
precession rate; geometry) rarely correspond to  astrophysical questions.  
We have demonstrated by example that measurements of relatively strong gravitational wave signals can  distinguish individual component masses
and spins to  astrophysical interesting accuracy [Fig. \ref{fig:Results:Intrinsic}].  Given the accuracy and 
number of measurements gravitational waves will provide, compared to existing astrophysical experience  \cite{PSconstraints3-MassDistributionMethods-NearbyUniverse,2010CQGra..27k4007M,2010ApJ...725.1918O,2012ApJ...757...55O,2013ApJ...778...66K}, these measurements should
transform our understanding of the lives and deaths of massive stars.   
Ignoring correlations, gravitational wave measurements seem to only relatively weakly constrain spin-orbit misalignent [Fig. \ref{fig:Results:Intrinsic}], a proxy for several
processes including supernova kicks and stellar dynamics.   
That said, gravitational wave measurements should strongly constrain the precession rate, a known expression of
spins, masses, and spin-orbit misalignment.    Formation models which make nontrivial predictions about both spin
magnitude and misalignment might therefore be put to a strong test with gravitational wave measurements.

\section*{Acknowledgements}
This material is based upon work supported by the National Science Foundation under Grant No. PHY-0970074, PHY-0923409, PHY-1126812,
and PHY-1307429.
ROS acknowledges support from the UWM Research Growth Initiative. 
BF is supported by an NSF fellowship DGE-0824162.
VR was supported by a Richard Chase Tolman fellowship at the California Institute of Technology.  
HSC, CK and CHL are supported in part by the
National Research Foundation Grant funded by the Korean
Government (No. NRF-2011-220-C00029) and the Global
Science Experimental Data Hub Center (GSDC) at KISTI.
HSC and CHL are supported in part by the BAERI
Nuclear R \& D program (No. M20808740002).
This work uses computing resources both at KISTI
and CIERA, the latter funded by NSF PHY-1126812.

\appendix
\section{More properties of the effective Fisher matrix}
\label{ap:RevisedEffectiveFisher}

\begin{table}
\begin{tabular}{l|l}
Type of joint information & Value \\\hline
$I(\{\mc,\eta,\chi_1,\beta_{JL}\}, \{\theta_{JN},\alpha_{JL}\}|\phi_{\rm ref})$ & 0.31 \\
$I( \{\theta_{JN}\},\{\alpha_{JL}\}|\{\mc,\eta,\chi_1,\beta_{JL},\phi_{\rm ref}\})$ & 0.01 \\
$I(\{\mc\}, \{\eta,\chi_1,\beta_{JL}\} | \{\theta_{JN},\alpha_{LN},\phi\})$ & 1.92
\end{tabular}
\caption{\label{tab:RevisedFisher:Information}\textbf{Separation of variables in the precessing effective Fisher matrix}: For several different subspaces
  $A,B$ and marginalized parameters $C$, the mutual
  information $I(A,B|C)$.
}
\end{table}
\subsection{Separation of scales and mutual information}
On physical grounds, we expect the timescales and modulations produced by precession to \emph{separate}, allowing
roughly independent measurements of orbital- and precession-rate-related parameters (i.e., $\mc,\eta,a, \beta_{JL}$) and purely geometric
parameters ($\alpha,\phi,\theta$).    
To assess this hypothesis quantitatively,  we evaluate the mutual information between the two subspaces.  For a
gaussian distribution described by a covariance matrix $\Gamma$, the mutual information between two subspaces $A,B$ is
[\abbrvAlignedPE{} Eq. (31)]:
\begin{eqnarray}
I(A,B) = -\frac{1}{2}\ln \frac{|\Gamma|}{|\Gamma_A||\Gamma_B|}
\end{eqnarray}
Table \ref{tab:RevisedFisher:Information} shows that after marginalizing out orbital phase ($\phi_{\rm ref}$), the mutual information between orbital-phase-related parameters
($\mc,\eta,a,\beta_{JL}$) and geometric parameters is small but nonzero (0.31): the two subspaces are weakly correlated.  
By comparison, the mutual information $I(a,c|B)$ between two intrinsic parameters $a,c$ in $A =\{\mc,\eta,\chi_1\}$ is large.
Finally,  after marginalizing out all other parameters, the mutual information between $\alpha_{JL}$ and
$(\theta_{JN})$ is small, as expected given the different forms in which these quantities enter into the outgoing
gravitational wave signal.

\subsection{Regularizing calculations with a prior}
Due to the wide range of eigenvalues and poor condition number,  all Fisher matrices are prone to numerical instability
in high dimension.   
Additionally, due to physical near-degeneracies, the error ellipsoid derived from the Fisher matrix alone may extend
significantly outside the prior range; see, e.g., examples in \cite{2013PhRvD..88h3002W}.  

\begin{widetext}
Following convention, to insure our results are stable to physical limitations, we derive parameter measurements accuracies $\Sigma =
\Gamma^{-1}/\rho^2$ by combining the signal amplitude $\rho$, the
normalized effective Fisher matrix $\hat{\Gamma}^{\rm eff}$ provided above, and  a prior $\Gamma^{\rm prior}$: 
\begin{eqnarray}
\Gamma^{\rm eff}_{\lambda} &\equiv \rho^2 \hat{\Gamma}^{\rm eff} + \lambda \Gamma^{\rm prior} \\
\Gamma^{\rm prior}          & = e_{\eta}\otimes e_{\eta} + \frac{e_a\otimes e_a}{20} +
  \frac{1}{(2\pi)^2}[
  e_\beta\otimes e_\beta+ 
  e_\theta\otimes e_\theta+ 
  e_\alpha\otimes e_\alpha+ 
  e_\phi\otimes e_\phi]
\end{eqnarray}
As expected given the eigenvalues and signal amplitude, this prior has no significant impact on our calculations.  In
particular, the eigenvalues and parameter measurement accuracies reported in the text are unchanged if this weak prior
is included.
\end{widetext}

\bibliography{paperexport}

\end{document}